\documentclass[aps,prl,twocolumn,
epsfig]{revtex4}
   
\usepackage{dcolumn}
\usepackage{bm}
\usepackage{graphicx}
\usepackage{amsmath}
\usepackage{latexsym}
\usepackage{amsfonts}
\usepackage{amssymb}
\usepackage{array}
\usepackage{epsfig}

\begin{document}

\title{Reply to Comment on ``Quantification of Macroscopic Quantum Superpositions within Phase Space"}
\author{Hyunseok Jeong, Minsu Kang and Chang-Woo Lee}

\affiliation{
Center for Macroscopic Quantum Control \& 
Department of Physics and Astronomy,
Seoul National University, Seoul, 151-747, Korea
}
\date{\today}

\maketitle

\setlength\arraycolsep{1pt}

Our recent Letter \cite{LeeJeong2011} proposes a general measure, $\cal I$,
to quantify macroscopic quantum superpositions. 
Gong points out \cite{Gong} a ``direct connection'' between
$\cal I$ and a previously studied quantity, 
$\chi^2$,
introduced as a measure of ``phase-space distributional heterogeneity'' 
\cite{Gu} to study 
classical and quantum chaos \cite{Gu,GongEtc}. 
He comments that $\cal I$ and $\chi^2$ are basically equivalent for pure states,
and ``closely related'' also for mixed states.

While it is 
beneficial to learn 
some unrecognized connection between 
the two measures, as we shall see,
$\cal I$ and $\chi^2$ are  
very different 
for some mixed states, and  $\chi^2$ does not work
as a sensible measure to quantify quantum superpositions.
In fact, $\chi^2$ has never 
been considered as a quantifier of quantum superpositions in the literatures \cite{Gu,GongEtc}.

According to the decohrence model considered in Ref.~\cite{LeeJeong2011},
${\cal I}=-\dot{\cal P}/2$ corresponds to ``purity decay rate'' while
$\chi^2=2(1-\dot{\cal P}/{\cal P})$ relates to ``purity decay rate divided by purity'',
where $\cal P$ and $\dot{\cal P}$ denote the purity and its time derivative, respectively.
In what follows, we present two specific examples to highlight the crucial difference
between the two measures.

\begin{figure}[b]
\includegraphics[width=200px]{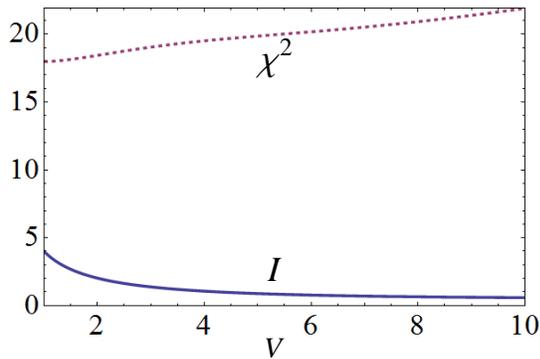}
\caption{Plots of $\mathcal{I}$ (solid curve) and $\chi^2$ (dotted curve) against $V$
for state $\rho_M$ with $d=1$.
}
\label{figure1}
\end{figure}

We first consider a superposition of thermal states introduced in Ref.~\cite{JeongRalph2006}:
\begin{equation}
\rho_M\propto \rho_{th}(V,d)+\rho_{th}(V,-d)+\sigma(V,d)+\sigma(V,-d)
\end{equation}
where 
$\rho_{th}(V,d)=\int d^2\alpha P_{th}(V,d)
|\alpha\rangle\langle\alpha|$ and
$\sigma(V,d)=\int d^2\alpha P_{th}(V,d)
|\alpha\rangle\langle -\alpha|$ 
with coherent state $|\alpha\rangle$ of amplitude $\alpha$ and
$P_{th}(V,d)=\frac{2}{\pi(V-1)}
\exp[-\frac{2|\alpha-d|^2}{V-1}]$.
Here, $V$ and $d$ respectively correspond to variance and
displacement of the displaced thermal state $\rho_{th}(V,d)$.
This type of state plays an important role in revealing some critical 
quantum behaviours 
 \cite{JeongRalph2006,JPR2009}. 
It is straightforward to obtain
$\cal I$ and $\chi^2$ for $\rho_M$ based on Refs.~\cite{LeeJeong2011} and \cite{Gu},
and the results for $d=1$ are plotted in Fig.~1.
Obviously, $\cal I$ and $\chi^2$ behave in opposite
ways: when $V$ increases,
$\cal I$ decreases to approach 1/2 \cite{correction} but $\chi^2$ keeps increasing without
limitation.

The second example further clarifies what happens. Let us consider a mixed state:
\begin{equation}
\rho_m=p|1\rangle\langle 1|+(1-p)\rho_{th}(V,0),
\end{equation}
where $|1\rangle$ is a single photon state and  
$V\rightarrow \infty$ is assumed \cite{note1}. 
We then observe 
 that $\chi^2$ of $\rho_m$ for an {\it arbitrarily} small value of $p$ 
approaches 6: this value of $\chi^2$ equals that of the single photon and far larger than that of
a thermal state $\chi^2\approx 0$ for $V\rightarrow \infty$.
When $p$ is very small, $\rho_m$ approaches a thermal state of $V\rightarrow \infty$
(or the identity operator), {\it i.e.}, states $\rho_m$ and
$\rho_{th}(V\rightarrow\infty,0)$ become nearly identical both theoretically and practically.
Therefore, such an asymptotic behaviour of $\chi^2$ does not
make sense as a measure of quantum superposition.
On the contrary, in the same limit, one can 
show that 
$\cal I$ of $\rho_m$ approaches zero, the same to that of $\rho_{th}(V\rightarrow\infty,0)$, 
which is intuitively acceptable.

The apparent difference between the two measures shown in the above examples is due to the  
fact that  
$\cal I$ relates to $\dot{\cal P}$ while $\chi^2$ to ${\dot{\cal P}}/{\cal P}$.
We conclude that the purity in the denominator of $\chi^2$ causes
this measure to {\it overestimate} the degree of superposition for
mixed states such as $\rho_M$ and $\rho_m$.

We finally note that in order to guarantee positivity of $\cal I$, one may simply 
remove factor $-1$ in definition (1) of our Letter \cite{LeeJeong2011}, and the
modified definition is equivalent to ${\cal I}=(P-{\dot P})/2$ in the relevant decoherence model.



\begin{thebibliography}{99}
\bibitem{note1}
We discuss an extreme case here, but qualitatively
the same conclusion is drawn from a finite $V\gg1$.
\end{thebibliography}
\end{document}